# Beyond Maintenance Manual Multimodal RAG: Suggesting What Tool

**Seongjun Ha**
**School of Aviation and Transportation Technology**
**Purdue University**
**1401 Aviation Drive**
**West Lafayette, IN, 47906**
**Ha92@purdue.edu**

**Md Rashedul Islam**
**Department of Industrial Engineering**
**Clemson University**
**Fernow St**
**Clemson, SC, 29634**
**mdrashi@clemson.edu**

*Abstract*—**Aircraft technicians are required to consult the maintenance manual (MM) for nearly every task, and locating the relevant procedure across hundreds of pages remains time-consuming. Multimodal retrieval augmented generation (MRAG) has been proposed to address this, allowing technicians to retrieve procedures, together with the accompanying figures, through natural-language queries. However, retrieval alone does not tell the technicians which tools the task requires. The MM identifies special tools only when the corresponding step is reached, and it does not state hand tool requirements at all; to select hand tools, technicians are required to find the hardware dimension from the illustrated parts catalog (IPC) and infer the right tool from it. We therefore propose MRAG-SWAT, an extension of the MRAG pipeline that returns the required hand tools and special tools alongside the retrieved procedure. The framework was implemented for the Lycoming IO-360-N1A engine and demonstrated on eight test queries. By presenting the correct tools together with the procedure, MRAG-SWAT may help reduce repeated trips to the tool crib, prevent damage to aircraft caused by improper tool selection, and thereby avoid additional maintenance tasks and support continued airworthiness.**

***Keywords—Multimodal RAG, Aircraft Maintenance, Tool Suggestions***

## I. Introduction

During aircraft maintenance, technicians are required to consult various maintenance documents to perform their tasks correctly and in the proper sequence. These documents include the aircraft maintenance manual (MM), illustrated parts catalog (IPC), service bulletins, airworthiness directives, and other related materials [1], [2]. Reviewing these documents is an essential step in every maintenance task; however, in practice, their poor usability makes maintenance tasks time-consuming and frustrating for technicians, because of the large volume of content and the difficulty of locating relevant pages [3], [4].

With recent advances in artificial intelligence, retrieval-augmented generation (RAG) models have been developed and adopted to retrieve aviation-related information through natural-language queries, demonstrating strong performance in this domain [5], [6], [7]. More recently, Ha et al. [8] proposed a multimodal RAG (MRAG) pipeline for the Cessna 172 MM, which incorporates multimodality into RAG to reflect the fact that technicians also rely on visual information during maintenance tasks. However, since these frameworks can only retrieve what has been stored in the vector database from the MM documents, we believe that MRAG for MM can be improved to better reflect technicians' actual workflow, thereby expanding its usability.

Specifically, in the context of aircraft maintenance processes, a prior study by Ha et al. [2] showed that technicians first review the assigned task to understand the required procedures and prepare the necessary tools before beginning the task. Yet the MM exhibits two key shortcomings in this regard. First, it does not clearly specify which hand tools are required to perform a given step [8]. Second, special tool requirements are often not indicated upfront but instead appear only when the corresponding step is reached in the procedure.

As a result, technicians estimate hardware sizes from experience, rely on their own technical judgment to select hand tools, and often make repeated trips to the tool crib for special tools after the task is already underway. These inefficiencies are most likely to occur when the overall procedure has not been reviewed in advance, or when steps are performed strictly in sequence without looking ahead. Furthermore, information about hardware size is not included in the MM; instead, it is found in the IPC. The MM focuses only on task procedures, while the IPC provides exploded views of systems along with component descriptions. Failure to use the correct tool may cause technicians to damage hardware or surrounding components, which in turn generates additional maintenance tasks.

These limitations stem both from the design of the MM and from existing RAG models, which retrieve only the information explicitly stored in their vector databases. Motivated by these gaps, we propose an advanced MRAG framework that additionally suggests the hand tools and special tools required for maintenance tasks.

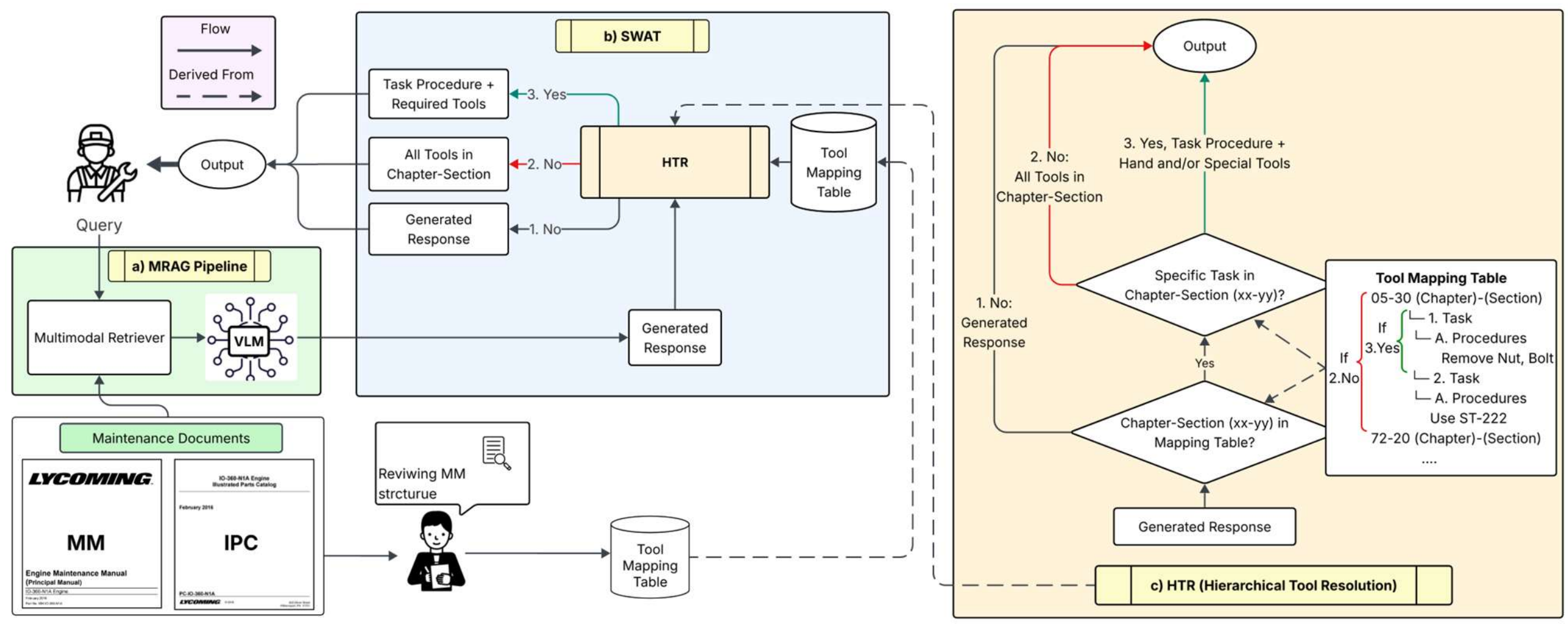

**Figure 1. MRAG-SWAT Framework: (a) MRAG pipeline, (b) SWAT, and (c) HTR**

## II. Research Objective

This research aims to develop a Multimodal Retrieval-Augmented Generation – Suggesting What Tool (MRAG-SWAT) framework that retrieves maintenance information while simultaneously suggesting the tools required based on the generated response.

## III. Methodology

To address the research objective, we developed the MRAG-SWAT framework. Figure 1 shows the overall process of the framework. MRAG-SWAT comprises two processes: the MRAG pipeline and SWAT. A technician submits a query to the MRAG pipeline, where it is processed by the ColPali retrieval operation [9]. Then, a vision-language model (VLM) processes the retrieved images and generates a response. The generated response is then passed to SWAT, which uses the tool mapping table and the hierarchical tool resolution (HTR) algorithm. SWAT then returns the suggested tools, so that the technician receives the task procedure together with the tools required to perform the query-related task. The two processes are described in the following sections:

### *A. MRAG Pipeline*

The MRAG pipeline of the MRAG-SWAT framework is illustrated in Figure 1 (a). The multimodal retrieval model uses the original ColPali [9] architecture, as in the MRAG pipeline of Ha et al. [8], which was validated on the Cessna 172 MM. The ColPali architecture retrieves image-based document pages, and the retrieved pages are then passed to the VLM, which generates a response to the query.

### *B. SWAT*

As shown in Figure 1 (b), the response generated by the MRAG pipeline is passed to SWAT, which produces tool suggestions based on a tool mapping table through the HTR algorithm. The details of the tool mapping table and HTR are described in the following sections:

#### *1) Tool Mapping Table*

To construct the tool mapping table, we reviewed how the MM specifies hardware and special tools and how it links to the IPC. We first reviewed the structure of the MM, which follows the ATA numbering format chapter-section (*xx-yy*), where *xx* denotes the chapter (system) and *yy* denotes the section (subsystem). Within a chapter-section, each task is given a specific name, and its procedures are listed beneath it. These procedures specify either a special tool or a hardware item required to perform the task, which form the two categories of tool requirements. For special tools, the MM gives the tool name and index number that are listed in the special tool table, which maps directly to that tool without an IPC lookup. For hardware such as a plug, screw, bolt, or nut, the MM instructs technicians to remove or install the item, which implicitly requires hand tools. Resolving a hardware item to a tool requires two attributes: the drive type of the item, which determines the type of tool, and its dimension, which determines the size of the tool. The MM names the hardware but does not specify these attributes, so each hardware item is required to be located in the corresponding parts table of the IPC, which lists the part number and a description containing the drive type and dimension. We linked hardware in the MM task procedures to the IPC in two ways: where the MM and the IPC share a common figure reference, the item number in the shared figure provides the match directly; where they do not, we searched the IPC manually to identify the matching hardware description. Once we identified all hardware items, we classified them by drive type to determine which hand tool applies, as given in Table 1, where nuts and bolts are grouped into a single category because both act on the same across-flats dimensions of threaded fasteners.

Table 1. Drive Type to Hand-Tool Category Mapping

| Head appearance | IPC hardware | Hand tool |
|---|---|---|
| ⬡ | Nut and Bolt | Socket and Wrench |
| □ | Square Head Plug | Wrench |
| ⊖ | Slotted Head Screw | Slotted Screwdriver |
| ⊕ | Phillips Head Screw | Phillips Screwdriver |
| ⬢ | Hex Socket Head Plug | Hex Key |

*Note. We assume the use of 12-point sockets, which are common in aviation maintenance and engage the six-point heads of the bolts and nuts listed above. Square head plugs are mapped to a wrench.*

To determine tool size from the hardware dimensions, we relied on published hardware sizing standards. Because hand tools act on standard hardware, the dimension-to-tool rules do not have to be aviation domain specific. Table 2 [10] lists the bolt diameter to socket/wrench size mapping; Table 3 [11] lists screw sizes for Phillips and slotted heads, converted to the nearest fractional equivalent using the size reference tables in [12], [13]; and Table 4 [14] lists hex-socket fittings such as NPT pipe plugs. The drive-type classification and the tool sizes obtained from this cross-document lookup were then recorded in the tool mapping table.

Table 2. Bolt/Nut Diameter to Socket/Wrench Size Mapping (in.)

| Bolt/Nut Diameter | Socket/Wrench Head Size | Bolt/Nut Diameter | Socket/Wrench Head Size |
|---|---|---|---|
| 1/4 | 7/16 | 9/16 | 13/16 |
| 5/16 | 1/2 | 5/8 | 15/16 |
| 3/8 | 9/16 | 3/4 | 1-1/8 |
| 7/16 | 5/8 | 7/8 | 1-5/16 |
| 1/2 | 3/4 | 1 | 1-1/2 |

Table 3. Screw Size to Phillips and Slotted Screwdriver Size (in.)

| Phillips Screwdriver | | | Slotted Screwdriver | | |
|---|---|---|---|---|---|
| Screw Size | Screw Diameter | Blade Size | Screw Size | Screw Diameter | Blade Size |
| 0 & 1 | 1/16–5/64 | #0 | 0 & 1 | 1/16–5/64 | 3/32 |
| 2-4 | 3/32–7/64 | #1 | 2 | 3/32 | 1/8 |
| 5-9 | 1/8–11/64 | #2 | 3 | 3/32 | 3/16 |
| 10-16 | 3/16–17/64 | #3 | 4-5 | 7/64–1/8 | 7/32 |
| 18-24 | 19/64–3/8 | #4 | 6-7 | 9/64–5/32 | 1/4 |
| | | | 8-10 | 11/64–3/16 | 5/16 |
| | | | 12-14 | 7/32-1/4 | 3/8 |
| | | | 16-18 | 17/64-19/64 | 7/16 |
| | | | 20-24 | 5/16–3/8 | 1/2 |

*Note. Phillips screwdriver blade sizes correspond directly to screw size. Slotted blade sizes are not standardized, so the slotted values are approximate.*

Table 4. Hex Socket Diameter to Hex Key Size Mapping (in.)

| NPT Plug Size | |
|---|---|
| Diameter | Hex Key Size |
| 1/16 | 5/32 |
| 1/8 | 3/16 |
| 1/4 | 1/4 |
| 3/8 | 5/16 |
| 1/2 | 3/8 |
| 3/4 | 9/16 |
| 1 | 5/8 |

### *2) Hierarchical Tool Resolution (HTR)*

Upon completion of the tool mapping table, we developed a hierarchical tool resolution (HTR) algorithm that produces tool suggestions by processing the generated response and referring to the tool mapping table, as shown in Figure 1 (c). The HTR algorithm is hierarchical in that it resolves a query at the two levels of the MM structure: the chapter-section level and the task level beneath it. The tool mapping table stores tools keyed to both levels, and the HTR algorithm decides which level to return based on how specific the generated response is.

Specifically, the HTR algorithm first inspects whether the generated response includes a chapter-section that is listed in the tool mapping table. If it does not, the generated response is returned unmodified, because the chapter-section has no associated hand or special tools. If it does, the HTR algorithm proceeds to the second decision, which checks whether the generated response identifies a specific task within that chapter-section. If no task is identified, all tools associated with the chapter-section are returned; that is, the algorithm collects the union of the tools mapped to every task under that chapter-section in the tool mapping table. If a task is identified, only the tools mapped to that task are returned, together with the task procedure. This hierarchical design reflects the fact that a technician may need tool information at either level: at the chapter-section level when preparing for a job, for example gathering every tool required for the task before accessing the aircraft, or at the task level when performing a specific procedure.

## IV. Case Study: Lycoming IO-360-N1A

We used the Lycoming IO-360-N1A engine as a case study, drawing on two maintenance documents for this engine: the IO-360-N1A engine MM [15] and its IPC [16]. The engine MM describes how to perform maintenance in terms of inspection, cleaning, repair, and assembly/disassembly, whereas the IPC provides an exploded view of the components together with a part list that gives an index number keyed to that view, the part number, and a description that includes size, dimensions, and quantity. The development of MRAG-SWAT for the IO-360-N1A is described in the following sections:

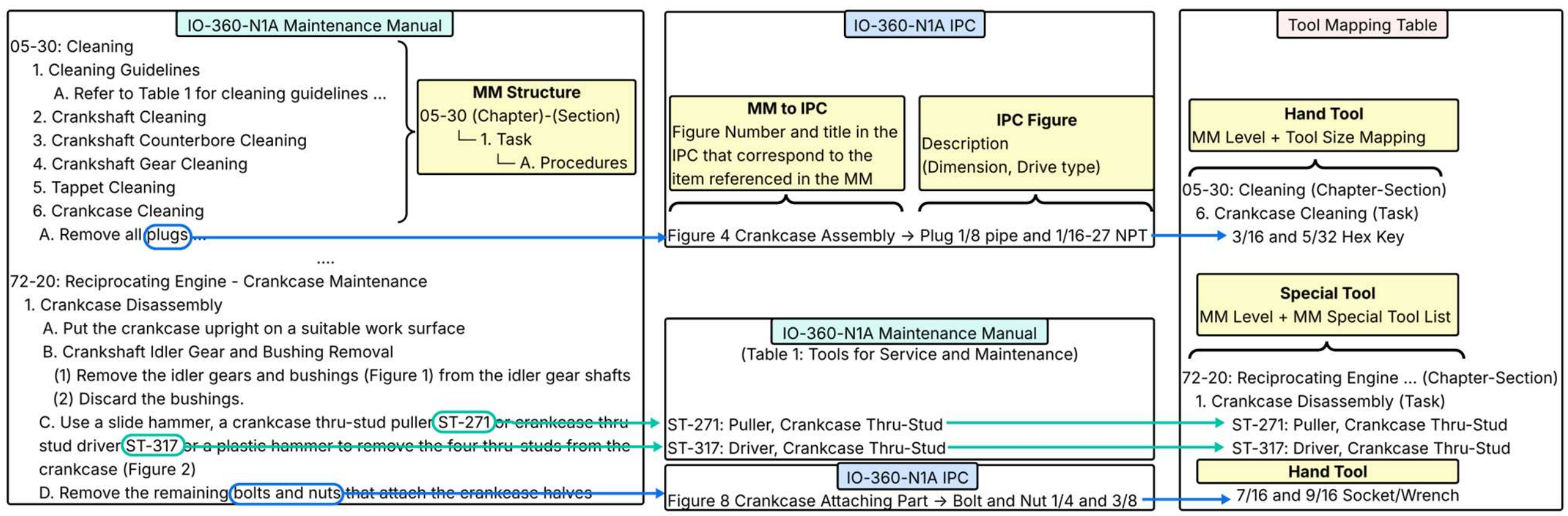

**Figure 2. Examples of Tool Mapping Table Construction for the Lycoming IO-360-N1A**

### A. *MRAG Pipeline for Lycoming IO-360-N1A*

We implemented the MRAG pipeline following Ha et al. [8], who applied the same architecture to the Cessna 172 MM using ColPali [9]. Among the available ColPali variants, we adopted ColQwen2, which uses a more advanced embedding backbone. The IO-360-N1A MM and IPC, a total of 279 PDF pages, were embedded with ColQwen2 on Google Colab with an H100 GPU. We then selected the GPT-4.1 model as the VLM that processes the retrieved images and generates a response to the query. Retrieval was configured to return the five most similar pages (top-$k$ = 5), following Ha et al. [8], who reported high retrieval performance at this setting.

### B. *SWAT for Lycoming IO-360-N1A*

To develop SWAT for the IO-360-N1A, we constructed a tool mapping table and an HTR algorithm by reviewing the MM and the IPC.

#### 1) *Tool Mapping Table for Lycoming IO-360-N1A*

Figure 2 shows the process of tool mapping table development for the Lycoming IO-360-N1A, which stores each hardware item with its dimension and the corresponding tool that HTR returns as a suggestion.

We reviewed the procedural sections of the IO-360-N1A MM, which follows the ATA chapter-section format. Each chapter-section is composed of numbered tasks, and the sublevels beneath each task describe its procedures, as illustrated in Figure 3:

Figure 3. Example of the Chapter-Section and Task Hierarchy in the IO-360-N1A MM

Cleaning (*05-30*) ← **Chapter-Section level**
  *1. Cleaning Guidelines* ← **Task Level**
    *A. Refer to Table 1 for cleaning guidelines for engine components.*
    *B. After the initial visual inspection….*
    *C. There are two processes for cleaning ….*
  *2. Crankshaft Cleaning*
    *A. Clean the inside of all crankpin journals, ….*
    *B. …*

Following the MM structure task by task, we identified the two categories of tool requirement: special tools, which the MM names directly, and hardware, whose removal or installation implies hand tools.

Specifically, for the special tools, the IO-360-N1A MM provides a special tool table in which each tool number is paired with its nomenclature and description, as shown in Table 5.

Table 5. Examples from the Special Tool Table in IO-360-N1A MM [15]

| Tool Number | Nomenclature and Description |
|---|---|
| ST-23 | Gage, Valve Clearance 0.028 to 0.080 in. |
| ST-25 | Compressor, Valve Spring |
| More … | |
| 64713 | Expander, Piston Ring |

To determine the hand tools, we reviewed the procedures of each task in the IO-360-N1A MM, as illustrated in Figure 3.

Each task lists every procedural step including the hardware to be removed or installed. Because some tasks in the MM and the IPC do not share a common figure, we matched each task to the IPC exploded-view figure covering its parts and noted the drive type and dimension of each hardware item from that figure's part number and description.

Based on the hardware identified from the IPC, we found four drive types in the IO-360-N1A: nut and bolt, square head plug, slotted head screw, and hex socket head plug. Nut and bolt form a single category, socket and wrench, because both act on the same across-flats dimension of threaded fasteners. Each drive type was mapped to its hand-tool category using Table 1, and the dimension obtained from the IPC was converted to a required tool size using Tables 2, 3, and 4.

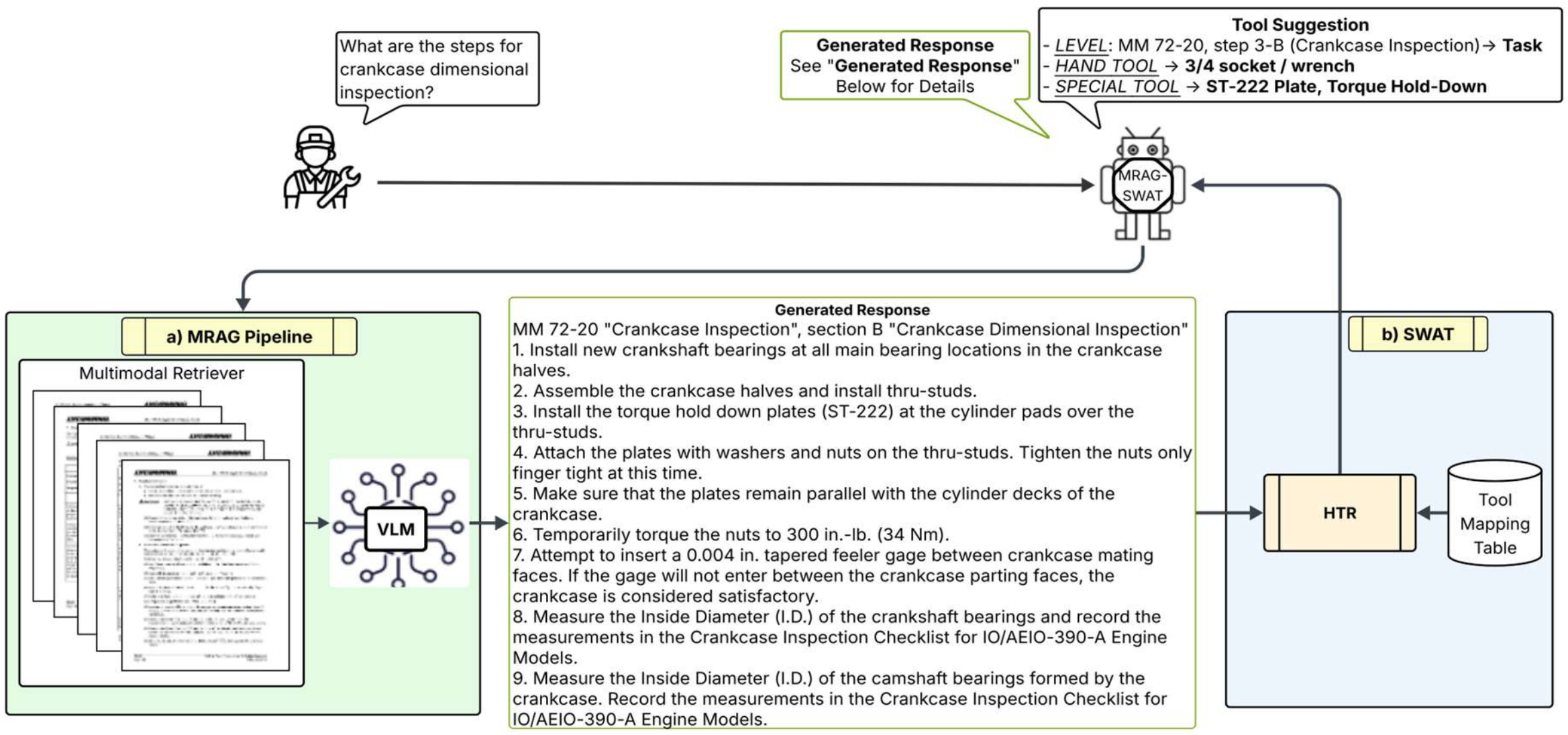

**Figure 4. MRAG-SWAT Validation Result for Query 8: Crankcase Dimensional Inspection**

*2) HTR for Lycoming IO-360-N1A*

The tool mapping table developed above is used by HTR, which processes the generated response in order to suggest tools. The HTR for the IO-360-N1A was developed based on the two levels of the MM structure. Figure 3 shows an example of this structure. When inspecting each response generated by the VLM, HTR makes the following two decisions in order:

1. Is the chapter-section of the generated response present in the IO-360-N1A tool mapping table?
2. If so, does the generated response include a specific task within that chapter-section?

Both decisions are made by matching the generated response against the chapter-section and task entries in the tool mapping table. The level of MM structure identified in the generated response determines the granularity of the suggestion: a query that identifies only a chapter-section returns all tools mapped to that section, whereas a task-specific query returns only the tools mapped to that task, together with the task procedure.

*C. Test Queries and Assessment*

To assess the operation of MRAG-SWAT, we constructed a test set of eight queries that cross two dimensions. The first dimension is the expected tool output: no tools, special tools only, hand tools only, or both, as determined by the tool mapping table. The second dimension is the tool suggestion level, reflecting the two HTR levels:

1. Chapter-section level: a general query for which SWAT returns all tools mapped to the chapter-section.
2. Task level: a specific query for which SWAT returns only the tools mapped to that task, together with the task procedure.

Each tool-output category was tested once at the chapter-section level and once at the task level, yielding the eight queries listed in Table 6. Each query was run through the MRAG-SWAT framework to produce a tool suggestion. Each suggested tool set was then assessed qualitatively against the ground truth derived from the MM and the IPC.

Table 6. Validation Query Scenarios and Corresponding Questions

| Query Scenarios | Query Questions | Query Level |
|---|---|---|
| 1. Query that does not require any tools | 1. Steps for unscheduled corrective maintenance | *N/A |
| | 2. Give me oil pressure relief valve inspection procedures | |
| 2. Special tools only | 3. What are the procedures for propeller flange bushing replacement? | 1 |
| | 4. Give me propeller flange bushing removal procedures | 2 |
| 3. Hand tools only | 5. Give me cleaning procedures | 1 |
| | 6. What are the procedures for cylinder removal? | 2 |
| 4. Both special and hand tools | 7. Give me the procedures for reciprocating engine – crankcase maintenance | 1 |
| | 8. What are the steps for crankcase dimensional inspection? | 2 |

** Tool suggestions are not applicable to query scenario 1, since it does not require any tools.*

## V. Results and Discussion

We evaluated MRAG-SWAT on the eight test queries. An example result is shown in Figure 4, and the complete set of results is provided in Appendix A. For the eight test queries, we first compared the HTR level classification by SWAT against the ground truth.

HTR classified six of the eight queries at the correct MM level. As shown in Appendix A, it misclassified the level for queries #5 and #7. For query #5, HTR classified the query at the task level (Cleaning Guidelines) although we expected the chapter-section level, so the hand tools mapped to Crankcase Cleaning in that chapter-section were not returned. For query #7, HTR classified the query at the task level although we expected the chapter-section level, so two special tools used in other tasks within that chapter-section were not returned.

We attribute this to the absence of a structured prompt template in MRAG response generation: because the generated response does not consistently state the chapter-section or the task, the level matched by SWAT depends on how consistently that information appears in the response, as the correct match for query #3 and the mismatches for queries #5 and #7 illustrate. A structured prompt template is therefore required to make level matching reliable, and further study is needed on prompt template design for response generation in MRAG and on its connection to SWAT.

For the tool suggestions, the suggested tools matched the ground truth in six of the eight queries; queries #5 and #7 returned incomplete tool sets. Specifically:

- Queries #1 and #2 contained no tool related information that HTR could match in tool mapping table, so no tools were suggested.
- Queries #3 and #4 returned special tools only.
- Query #6 returned hand tools only.
- Query #8 returned both special tools and hand tools.

In addition, we believe that the MRAG-SWAT framework has potential beyond tool suggestion: because the tool mapping table already links each task to its IPC entries, the same mapping can also return part numbers and related hardware information, thereby improving usability for technicians.

However, MRAG-SWAT also inherits a limitation from the ColPali architecture. ColPali retrieves on semantic similarity alone and therefore treats each page as a self-contained unit, so when a procedure continues onto the following page, the continuation is not retrieved with it. This behavior appeared in query #6, in which MRAG did not retrieve the full procedure but SWAT suggested hand tools. This is because SWAT resolves the query level from the generated response and matches it against the tool mapping table, which returns either all tools for the chapter-section or the tools for the specific task.

One remedy for the limitation above is to redesign the MM itself, as Ha et al. [8] suggested in their discussion. A related obstacle emerged in our case study: in the Lycoming IO-360-N1A documentation, the MM and the IPC do not share a common table of contents, and figure names in the manual do not correspond to figure names in the IPC. Because the direct figure–item path from the MM into the IPC is unavailable, each hardware cue had to be matched by chapter and component instead, and part-number resolution fell back on manual interpretation.

## VI. Limitations

We also identified two limitations of this research:

1. Technicians frequently use extensions, adapters, or offset drives to reach components in confined locations, and the required tools may vary depending on accessibility and fastener dimension. Since SWAT maps dimensions to tools deterministically and does not consider the physical task situation, such accessibility related factors fall outside its current scope.
2. We made the best effort to match hardware dimensions to the correct tools, but some specifications remain ambiguous; fillister head slotted screws, for example, specify head style but not slot width. The tool mapping table therefore requires review by aircraft maintenance experts before it is deployed to maintenance personnel.

## VII. Conclusion

Consulting the MM is essential for technicians to properly perform maintenance, but searching the manual for the correct information is time-consuming, and the correct page is often hard to locate. While RAG models have achieved high performance in retrieving MM information in previous studies, aircraft MMs do not state what tools are needed for maintenance, which may lead technicians to use an improper tool. Therefore, we developed the MRAG-SWAT framework, which extends the usability of conventional MRAG by adding a feature that suggests tools linked to the IPC. We believe this framework can help technicians obtain MRAG responses together with tool suggestions, enabling them to properly perform maintenance tasks with the correct tools.

## Acknowledgements

This research did not receive any specific grant from funding agencies in the public, commercial, or not for profit sectors.

## Declaration of Generative AI in Scientific Writing

During the preparation of this work the author(s) used Microsoft Copilot and Claude Sonnet 5 to refine grammar and improve the flow of the text. After using this tool/service, the author(s) reviewed and edited the content as needed and take(s) full responsibility for the content of the published article.

# Appendix A: MRAG-SWAT Outputs and Ground Truth for the Eight Test Queries

| Query Level | Query | MRAG-SWAT Response | | GT |
|---|---|---|---|---|
| | | MRAG | SWAT | Tool Suggestion |
| *N/A | 1. Steps for unscheduled corrective maintenance | Unscheduled corrective maintenance is necessary when any of the following conditions occur:<br>Lightning strike, Engine overspeed, Incorrect fuel or fuel contamination, Soaked engine, Engine on fire or near fire Hydraulic lock, Volcanic ash/particulate contamination, Valve sticking, Oil starvation/sudden loss of oil pressure | - *LEVEL*: MM 05-50, (Unscheduled Corrective Maintenance) → **Chapter-Section**<br>- *HAND TOOL:* **None**<br>- *SPECIAL TOOL*: **None** | **-** *LEVEL*: **Chapter-Section**<br>- *HAND TOOL*: **None**<br>- *SPECIAL TOOL*: **None** |
| | 2. Give me oil pressure relief valve inspection procedures | MM 72-50 "Oil Pressure Relief Valve Inspection", step 5<br>1. Examine the oil pressure relief valve spring to be sure it meets specifications in accordance with the latest revision of Service Table of Limits - SSP-1776.<br>2. Examine the threads on the oil pressure relief valve. If the threads are stripped or galled, replace the valve. | - *LEVEL*: MM 72-50, step 5 (Oil Pressure Relief Valve Inspection) → **Task**<br>- *HAND TOOL*: **None**<br>- *SPECIAL TOOL*: **None** | - *LEVEL*: **Task**<br>- *HAND TOOL*: **None**<br>- *SPECIAL TOOL*: **None** |
| 1 | 3. What is the procedure for propeller flange bushing replacement? | MM 72-15 "Propeller Flange Bushing Replacement"<br>1. Propeller Flange Bushing Removal<br>A. During removal of the bushings from the propeller flange (also known as the crankshaft flange), attach a removable non-adhesive label/tag on the bushing that identifies the correct bushing part number and location for reference on reassembly.<br>B. Refer to the latest revision of Service Instruction No. SI-1098 for the latest part number and propeller flange bushing location.<br>C. Use the Propeller Flange Bushing Removal/Installation Tool ST-115 to remove each bushing from the propeller flange.<br>2. Propeller Flange Bushing Installation<br>A. As per the latest revisions of Service Instruction No. SI-1098, identify the correct bushing part numbers for your engine and the location of each bushing.<br>B. Use the Propeller Flange Bushing Removal/Installation Tool ST-115 to install the bushings of the correct part number on the propeller flange on the crankshaft in the location designated for your engine model specified in the latest revision of Service Instruction No. SI-1098. | - *LEVEL:* MM 72-15 (Propeller Flange Bushing Replacement) → **Chapter-Section**<br>- *HAND TOOL:* **None**<br>- *SPECIAL TOOL* > **ST-115 Tool, Install and Remove Propeller Flange Bushings** | - *LEVEL*: **Chapter-Section**<br>- *HAND TOOL*: **None**<br>- *SPECIAL TOOL* > **ST-115** |
| 2 | 4. Give me propeller flange bushing removal procedures | MM 72-15 "Propeller Flange Bushing Replacement", step 1 "Propeller Flange Bushing Removal"<br>1. During removal of the bushings from the propeller flange (also known as the crankshaft flange), attach a removable non-adhesive label/tag on the bushing that identifies the correct bushing part number and location for reference on reassembly.<br>2. Refer to the latest revision of Service Instruction No. SI-1098 for the latest part number and propeller flange bushing location (since an update of Service Instruction No. SI-1098 is more likely to occur before the parts catalog update).<br>3. Use the Propeller Flange Bushing Removal/Installation Tool ST-115 to remove each bushing from the propeller flange. | - *LEVEL*: MM 72-15, step 1 (Propeller Flange Bushing Removal) → **Task**<br>- HAND TOOL: **None**<br>- *SPECIAL TOOL* > **ST-115 Tool, Install and Remove Propeller Flange Bushings** | - *LEVEL*: **Task**<br>- *HAND TOOL*: **None**<br>- *SPECIAL TOOL* > **ST-115 Tool** |
| 1 | 5. Give me cleaning procedures | MM 05-30 "CLEANING", step 1 "Cleaning Guidelines"<br>A. Refer to Table 1 for cleaning guidelines for engine components.<br>B. After the initial visual inspection (in Chapter 05-20 of the IO-360-N1A Engine Service Manual), clean engine parts thoroughly per instructions in this chapter.<br>(1) Degreasing removes dirt and sludge (soft carbon). Soak the component or part in mineral spirits or other degreaser.<br>(2) Decarbonizing removes hard carbon with an initial soak of the part in a warm or heated decarbonizing solution. After the soak, use a (non-wire) bristle brush, wooden scraper, or grit-blasting (with non-abrasive media) to physically remove the hard carbon. | - *LEVEL*: MM 05-30; step 1 (Cleaning Guidelines) → **Task**<br>- *HAND TOOL:* **None**<br>- *SPECIAL TOOL*: **None** | - *LEVEL*: **Chapter-Section**<br>- *HAND TOOL:*<br>> **3/16 hex key**<br>> **5/32 hex key**<br>- *SPECIAL TOOL*: **None** |
| 2 | 6. What are the procedures for cylinder removal? | MM 72-30 "Cylinder Removal", steps A–H "Cylinder Removal"<br>A. Disable all power to the engine and disconnect the engine wiring harness from the airframe.<br>B. If not already done:<br>Disconnect the ignition leads to the spark plugs. Remove the top and bottom spark plugs from the cylinder.<br>Remove the intake and exhaust pipes from the cylinder to be removed.<br>Remove the clamps that attach the fuel line to the shroud tube, disconnect the fuel line from the injection nozzle, remove the injection nozzle from the cylinder, and reconnect the removed injection nozzle to the fuel line.<br>C. Oil Drain Tube Removal<br>(1) Remove and discard the hose clamps.<br>(2) Disconnect the oil drain tube fitting from the engine cylinder.<br>(3) Remove and discard the hose. | - *LEVEL*: MM 72-30, step A–H (Cylinder Removal) → **Task**<br>- *HAND TOOL*<br>> **3/4 socket / wrench**<br>> **3/8 blade slotted screwdriver**<br>> **9/16 socket / wrench**<br>- *SPECIAL TOOL*: **None** | - *LEVEL*: **Task**<br>- *HAND TOOL*<br>> **3/4 socket / wrench**<br>> **3/8 blade slotted screwdriver**<br>> **9/16 socket / wrench**<br>- *SPECIAL TOOL*: **None** |

| 1 | 7. | Give me the procedures for reciprocating engine – crankcase maintenance | MM 72-20 "Reciprocating Engine – Crankcase Maintenance", step 1 "Crankcase Disassembly"<br>1. Put the crankcase upright on a suitable work surface.<br>2. Remove the idler gears and bushings from the idler gear shafts. Discard the bushings.<br>3. Remove all thru-studs, **nuts**, and **bolts** from the crankcase halves before separating the crankcase halves to prevent damage.<br>4. Use a slide hammer, crankcase thru-stud puller **ST-271** or crankcase thru stud driver **ST-317**, or a plastic hammer to remove the four thru-studs from the crankcase.<br>5. Remove the remaining **bolts** and **nuts** that attach the crankcase halves.<br>6. Insert one used pushrod into each of the four holes where the thru-studs were removed to support the camshaft and crankshaft when the crankcase halves are separated.<br>MM 72-20 "Crankcase Inspection", step 3 "Crankcase Inspection"<br>1. Examine the exterior surface of the crankcase for cracks and damage. Replace a crankcase with one or more cracks.<br>2. Complete a check of the torque on the crankcase hardware fasteners per the latest revision of Service Table of Limits - SSP-1776. | - *LEVEL*: MM 72-20, step 1 (Crankcase Disassembly) → **Task**<br>- *HAND TOOL*<br>**> 1/2 socket / wrench**<br>**> 3/16 hex key**<br>**> 3/4 socket / wrench**<br>**> 3/8 hex key**<br>**> 5/16 blade slotted screwdriver**<br>**> 5/16 hex key**<br>**> 5/32 hex key**<br>**> 7/16 socket / wrench**<br>**> 9/16 socket / wrench**<br><br>- *SPECIAL TOOL* :<br>> **ST-271 Puller, Crankcase Thru-Stud**<br>> **ST-317 Driver, Crankcase Thru-Stud**<br>> **ST-389 Tool, Crankcase Separating** | - *LEVEL*: **Chapter-Section**<br><br>- *HAND TOOL*<br>**> 1/2 socket / wrench**<br>**> 3/16 hex key**<br>**> 3/4 socket / wrench**<br>**> 3/8 hex key**<br>**> 5/16 hex key**<br>**> 5/32 hex key**<br>**> 7/16 socket / wrench**<br>**> 9/16 socket / wrench**<br>**> 5/16 blade slotted screwdriver**<br><br>- *SPECIAL TOOL* :<br>**ST-222 ←Not returned by SWAT**<br>**ST-271**<br>**ST-317**<br>**ST-383 ←Not returned by SWAT**<br>**ST-389** |
|---|---|---|---|---|---|
| 2 | 8. | What are the steps for crankcase dimensional inspection? | MM 72-20 "Crankcase Inspection", section B "Crankcase Dimensional Inspection"<br>1. Install new crankshaft bearings at all main bearing locations in the crankcase halves.<br>2. Assemble the crankcase halves and install thru-studs.<br>3. Install the torque hold down plates (**ST-222**) at the cylinder pads over the thru-studs.<br>4. Attach the plates with washers and **nuts** on the thru-studs. Tighten the **nuts** only finger tight at this time.<br>5. Make sure that the plates remain parallel with the cylinder decks of the crankcase.<br>6. Temporarily torque the **nuts** to 300 in.-lb. (34 Nm).<br>7. Attempt to insert a 0.004 in. tapered feeler gage between the crankcase mating faces. If the gage will not enter between the crankcase parting faces, the crankcase is considered satisfactory.<br>8. Measure the Inside Diameter (I.D.) of the crankshaft bearings and record the measurements in the Crankcase Inspection Checklist for IO/AEIO-390-A Engine Models.<br>9. Measure the Inside Diameter (I.D.) of the camshaft bearings formed by the crankcase. Record the measurements in the Crankcase Inspection Checklist for IO/AEIO-390-A Engine Models. | - *LEVEL*: MM 72-20, step 3-B (Crankcase Inspection) → **Task**<br>- *HAND TOOL* > **3/4 socket / wrench**<br>- *SPECIAL TOOL* > **ST-222 Plate, Torque Hold-Down** | - *LEVEL*: Task<br>- *HAND TOOL* > **3/4 socket / wrench**<br>- *SPECIAL TOOL* > **ST-222 Plate** |